# EFFICIENT OFFLINE ALGORITHMIC TECHNIQUES FOR SEVERAL PACKET ROUTING PROBLEMS IN DISTRIBUTED SYSTEMS

Mugurel Ionuţ Andreica, Nicolae Ţăpuş

Abstract. In this paper we consider several problems concerning packet routing in distributed systems. Each problem is formulated using terms from Graph Theory and for each problem we present efficient, novel, algorithmic techniques for computing optimal solutions. We address topics like: bottleneck paths (trees), optimal paths with non-linear costs, optimal paths with multiple optimization objectives, maintaining aggregate connectivity information under a sequence of network link failures, and several others.

2000 *Mathematics Subject Classification*: 90B18, 90B20, 90B25, 90B35, 90C27, 90C35, 90C39, 90C47, 94A05.

## 1. Introduction

The development of distributed systems worldwide follows a steeper and steeper ascending trend, as such systems become natural solutions to many real-life problems. Communication (at a lower level) and collaboration (at a higher level) are two key elements of a distributed system. At the lowest level, packet routing techniques are employed by the intermediate nodes, switches and routers in order to transfer packets from a source to one or several destinations. However, we have little control over the packet routing techniques employed by the routers in the Internet and, furthermore, the Internet does not provide any kind of Quality-of-Service (QoS) guarantees. Because of this, devising new communication architectures and novel, efficient packet routing algorithms and techniques is an important step towards obtaining increased communication performance and QoS guarantees.

In this paper we present novel algorithmic techniques for several offline packet routing problems. The offline property means that all the required information is stable and available in advance. Moreover, it is assumed that we have full control over the entire distributed system and its topology. These assumptions are somewhat unrealistic for a practical setting if we consider the entire distributed system. However, they are useful as a theoretical basis for evaluating the performance of online techniques, and as practical tools, if we restrict their scope to limited parts of a large distributed systems.

The rest of this paper is structured as follows. In Sections $2-5$ we discuss several problems regarding bottleneck paths (trees). We present (nearly) optimal algorithms both for computing only one such path (tree) and for answering efficiently multiple path computation queries. In Sections $6-9$ we consider several optimization problems regarding optimal paths with non-linear costs, ordering constraints, or multiple objectives. In Section 10 we consider the problem of maintaining aggregate connectivity information under a sequence of edge deletions in an undirected network. In Section 11 we present related work and in Section 12 we conclude.

## 2. Maximum Capacity Path (Tree)

We are given a directed graph with $n$ vertices and $m$ edges. Each directed edge $(u,v)$ has a capacity $c(u,v) \geq 0$. In the *Unconstrained Maximum Capacity Path* problem we need to find a path having maximum capacity between a pair of vertices $s$ and $t$. A path is a sequence of vertices $v(1), v(2), \ldots, v(q)$ ($q > 0$), where $v(1) = s, v(q) = t$ and there exists a directed edge $(v(i), v(i+1))$ ($1 \leq i \leq q-1$). The capacity of the path is the minimum capacity of the edges composing the path, i.e. $min\{c(v(i), v(i+1)) | 1 \leq i \leq q-1\}$. In order to find a maximum capacity path from $s$ to $t$, we can use a modified Dijkstra's algorithm. We will compute $cmax(i)$=the maximum capacity of a path from $s$ to $i$ and, when expanding a vertex $i$, we consider every directed edge $(i,j)$ and update $cmax(j)$ to $max\{cmax(j), min\{cmax(i), c(i,j)\}\}$ (initially, we have $cmax(s) = +\infty$ and $cmax(i) = 0$, for $i \neq s$). $cmax(t)$ contains the maximum capacity of a path from $s$ to $t$. By storing "parent" pointers ($parent(j) = i$, if $i$ was the last vertex whose expansion updated $cmax(j)$), we can reconstruct the actual path (from $s$ to $t$). The time complexity of this approach is $O(m \cdot log(n))$ (when implementing Dijkstra's algorithm using a priority queue) or $O(n^2)$. If the capacities are integer numbers bounded by a small constant $CAPMAX$, the time complexity can become $O(n+m+CAPMAX)$, by maintaining a linked-list $LL(c)$ for every possible value of the capacity $c$ ($0 \leq c \leq CAPMAX$). Initially, $LL(CAPMAX)$ contains only the vertex $s$ (or, if we start from multiple sources, $LL(CAPMAX)$ contains all of these sources) and $LL(0)$ contains all the other vertices. We traverse the capacities from $CAPMAX$ down to 0 and for each capacity $c$ we traverse the linked-list $LL(c)$ and expand every vertex $i$ in it (this means that $cmax(i) = c$). When expanding a vertex $i$, we consider all the edges $(i, j)$ and update $cmax(j)$ (as before). If the value $cmax(j)$ changes, we remove vertex $j$ from its old linked-list and insert it into $LL(cmax(j))$ (it is possible even that $cmax(j) = c$). A variation of this technique uses *lazy deletion*, i.e. it does not delete the vertex $j$ from its old linked-list and just copies it to the new linked-list (corresponding to a higher capacity). Then, when traversing the vertices $i \in LL(c)$, we also need to verify if $cmax(i) = c$; if it is not, then this occurrence of the vertex $i$ is "old" and will be ignored. Another approach is based on binary searching the capacity $Cpath$ of the path from $s$ to $t$. We choose a candidate capacity $Cpath$ and then we perform a feasibility test, in order to verify if a path with capacity larger than or equal to $Cpath$ exists from $s$ to $t$. The feasibility test runs in $O(n+m)$ time. All the edges with capacities smaller than $Cpath$ are ignored and then we perform a depth-first or breadth-first search (DFS or BFS) starting from $s$. If vertex $t$ is visited during this search, then a path with capacity $c \geq Cpath$ exists and we can test a larger value of $Cpath$; otherwise, we test a smaller value. The time complexity of this approach is $O((n+m) \cdot log(m))$ if we sort all the edges initially (according to their capacities) and then we choose the value $Cpath$ from the set of edge capacities, or $O((n+m) \cdot log(CAPMAX))$ if we binary search the capacity in the interval $[0, CAPMAX]$, where $CAPMAX$ is the maximum capacity of an edge (in this case, if the capacities are not integers, we stop the binary search when the search interval becomes smaller than a constant $\varepsilon > 0$). The algorithm also works for undirected graphs, by transforming them into directed graphs: every undirected edge $(u, v)$ is replaced by two directed edges $(u, v)$ and $(v, u)$ with the same capacities as the undirected edge.

The *Unconstrained Maximum Capacity Multicast Tree* problem asks for a directed multicast tree from a source vertex $s$ to a subset of vertices $D=\{d(1), d(2), \ldots, d(k)\}$ (called *destinations*) with maximum capacity. The capacity of a tree is the minimum capacity of an edge of the tree. The tree may contain additional vertices (except the source vertex $s$ and the destinations). We can use the same techniques we used for the *Unconstrained Maximum*

*Capacity Path* problem. By computing the same values $cmax(i)$ for each vertex $i$, representing the maximum capacity of a path from $s$ to $i$, the maximum capacity of a multicast tree is $min\{cmax(d(j))|1 \leq j \leq k\}$. The actual tree is obtained as the union of the paths from the source vertex $s$ to every destination $d(j)$ ($1 \leq j \leq k$). We consider the destinations in some order and then follow the "parent" pointers we stored during the computation of the optimal paths, all the way to the source vertex $s$. This approach constructs the tree in $O(n \cdot k)$ time, as the length of every path from $s$ to a destination may contain $O(n)$ vertices. In order to reduce the complexity of the tree construction phase to $O(n)$, we will mark every vertex visited by following the "parent" pointers. When we follow these pointers starting from a destination $d(j)$, we stop when we reach the source vertex $s$ or when we reach a vertex which has previously been marked. The parent in the tree of every marked vertex $v$ (except $s$) is its predecessor (i.e. "parent") on the optimal path from $s$ to $v$.

We can also use binary search to look for the capacity $Ctree$ of the tree. Then, we perform a feasibility test, in order to check if a tree with capacity larger than or equal to $Ctree$ exists. We ignore all the edges with capacity smaller than $Ctree$ and perform a BFS or DFS traversal from the vertex $s$. If all the destinations are reachable, then such a tree exists and we will test a larger value of $Ctree$ next; otherwise, we test a smaller value. After finding the optimal capacity and obtaining a DFS (or BFS) tree rooted at $s$, we repeatedly remove a leaf from this tree, if the leaf is not one of the destinations. We can perform this stage in $O(n)$ time by recursively traversing the tree, starting from the root; when we return from the recursive traversals of all of vertex $v$'s sons and $v$ has no (more) sons, we check if $v$ is one of the destinations; if it is not, then we remove $v$ from the tree and remove it from the list of sons of its parent in the tree (or we simply decrement by 1 the number of actual sons vertex $v$'s parent still has). Both algorithms also work for undirected graphs, using the same transformation described previously.

## 3. MAXIMUM CAPACITY PATH QUERIES

We are given a connected, undirected graph composed of $n$ vertices and $m$ edges. Each edge $(u, v)$ has a capacity $cap(u, v)$. We want to be able to answer the following types of queries efficiently: what is the maximum capacity of a path between two vertices $u$ and $v$ ? The capacity of a path between two vertices $u$ and $v$ is the minimum capacity of an edge on the path. We could use one of the algorithms described in the previous section in order to find the answer for each query, but this would be too inefficient. Instead, we will preprocess the graph into a data structure using a solution for the *Union-Find* problem [4], such that each query can be answered in $O(log(n))$ time. We first sort all the edges of the graph in decreasing (non-increasing) order of their capacities. Then, we construct $n$ disjoint sets, one for every vertex of the graph and we start traversing the edges in the sorted order. When encountering an edge $(u, v)$, we compute $ru = Find(u)$ and $rv = Find(v)$, the representatives of the sets of the vertices $u$ and $v$. If $ru \neq rv$, then we will unite the sets corresponding to $ru$ and $rv$. We do this using the *union by rank* heuristic and setting $parent(ru) = rv$, if the height of the subtree rooted at $ru$ is smaller than the height of the subtree rooted at $rv$ (otherwise, we set $parent(rv) = ru$). Assuming that we are in the case $parent(ru) = rv$, we update the height of the subtree rooted at $rv$ (as $max\{h_{(old)}(rv), h(ru) + 1\}$) and set the weight of the tree edge $(rv, ru)$ to $cap(u, v)$; the other case is symmetrical. We will not use the path compression technique in combination with the union by rank heuristic (as it is customary). This way, once an edge $(rv, ru)$ is added to a tree, it will continue to remain a tree edge. We now add a special vertex $r$ and connect $r$ to the root vertex $ru$ of every tree (representing a connected component) by an edge of capacity 0 (and set $parent(ru) = r$).

In order to answer maximum capacity path queries, we perform a DFS traversal from $r$ and compute the level of every vertex in the tree. We have $level(r) = 0$ and $level(u) = level(parent(u)) + 1$. Then, for each query asking for the largest capacity of a path between two vertices $u$ and $v$, we perform the following actions. We initialize $pu$ to $u$ and $pv$ to $v$. Then, as long as $pu \neq pv$, we set $pu$ to $parent(pu)$, if $level(pu) > level(pv)$ (otherwise, we set $pv = parent(pv)$). The maximum capacity of a path between $u$ and $v$ is the minimum capacity of an edge on the path between $u$ and the final value of $pu$ or on the path between $v$ and the final value of $pv$. Because using the union by rank heuristic the height of every tree is $O(log(n))$, it takes $O(log(n))$ steps before $pu$ and $pv$ become equal to $LCA(u, v)$ (the lowest common ancestor of the vertices $u$ and $v$) and there are $O(log(n))$ edges on the paths between $u$ and $LCA(u, v)$, and $v$ and $LCA(u, v)$. We will name this simple technique of computing the LCA of two vertices the *level-by-level technique*.

The preprocessing stage takes $O(m \cdot log(m) + m \cdot log(n) + n)$ time and each query takes $O(log(n))$ time. We can improve the query time in several ways, by performing some extra preprocessing. Let's assume that the height of the tree is $H$. We perform a DFS traversal of the tree, starting from the root. During this traversal, we perform several actions. Each vertex $i$ is assigned its corresponding DFS number, $DFSnum(i)$ ($DFSnum(i) = j$ if vertex $i$ is the $j^{th}$ different vertex visited during the tree traversal); for each vertex $i$ we compute $DFSmax(i)$, the maximum DFS number of a vertex in its subtree. $DFSmax(i)=\max\{DFSnum(i), DFSmax(s(i,1)), \ldots, DFSmax(s(i,ns(i)))\}$ ($ns(i)$=the number of sons of vertex $i$ and $s(i,j)$=the $j^{th}$ son of vertex $i$). During the traversal we maintain a stack $S$ of the visited vertices (the vertices on the path from the root to the current vertex). We denote by $S(i)$ the $i^{th}$ entry in the stack, counting from the bottom ($S(1)$ is the tree root). Let's assume that $S(top)$ is the currently visited vertex. Then $S(top-j)$ is the $j^{th}$ ancestor of $S(top)$. For each vertex $i$ we store two values: $Anc(i,1)$=the parent of the current vertex $i$ (i.e. $S(top-1)$); $Anc(i,2)$=the ancestor of vertex $i$ which is located $k$ levels above (i.e. $S(top-k)$, or $S(1)$ if $top-k \leq 0$). $k$ is a value which depends on $H$ (we will show that $k = H^{\frac{1}{2}}$ is a good choice). We will use the values $Anc(i,*)$ in order to compute efficiently the lowest common ancestor of any two vertices $u$ and $v$. We will initialize a pointer $pu$ to $u$. Then, while $DFSnum(v) \notin [DFSnum(pu), DFSmax(pu)]$, we set $pu = Anc(pu, 2)$. Eventually, we will reach a vertex $pu$ (possibly the tree root) where the condition holds. This vertex is an ancestor of $LCA(u,v)$. Let's denote by $ppu$ the previous value of $pu$ (the one before reaching the final value). Obviously, we have $Anc(ppu, 2) = pu$. We will now move $ppu$ up the tree level by level. As long as $DFSnum(v) \notin [DFSnum(ppu), DFSmax(ppu)]$, we set $ppu = Anc(ppu, 1)$. The final value of $ppu$ is $LCA(u,v)$. The time complexity of this approach is $O(k + H/k)$ per query. By choosing $k = H^{\frac{1}{2}}$, we obtain a very practical $O(H^{\frac{1}{2}})$ algorithm for computing the lowest common ancestor of any two vertices (using $O(n)$ preprocessing and $O(n)$ memory storage). There are better algorithms for LCA queries for any pair of vertices in a tree, but the one mentioned before is extremely easy to implement. For instance, there exists an algorithm with $O(n)$ preprocessing and $O(1)$ LCA query time (but it is quite difficult to implement) [5] and another one, based on the jump pointers method, with $O(n \cdot log(n))$ preprocessing time and $O(log(n))$ query time [10]. After computing $LCA(u,v)$, we have two cases. If $LCA(u,v) = u$, then the answer to the query is the capacity of the last tree edge on the path between $v$ and $u$ (the edge adjacent to vertex $u$ on this path). A similar argument holds for the case $LCA(u,v) = v$. In the second case, $LCA(u,v) \neq u$ and $LCA(u,v) \neq v$. The answer is the minimum of the capacities of the edges $(su, LCA(u,v)), (sv, LCA(u,v))$, where $su$ ($sv$) is the son of $LCA(u,v)$ con-taining $u$ ($v$) in its subtree. In order to compute $su$ ($sv$), we start from $u$ ($v$) and advance along the $Anc(*, 2)$ pointers, until we reach the highest ancestor $a$ whose interval $[DFSnum(a), DFSmax(a)]$ is strictly included inside $[DFSnum(LCA(u, v)), DFSmax(LCA(u, v))]$ Afterwards, we advance along the $Anc(*, 1)$ pointers until we reach the highest ancestor with the same property; this ancestor is $su$ ($sv$).

## 4. Farthest Distance Path (Tree)

We are given an undirected graph with $n$ vertices and $m$ edges. Each edge $(u, v)$ has a length $len(u, v) \geq 0$. We are also given $k \leq n$ obnoxious vertices $ov(1)$, ..., $ov(k)$ (e.g. vertices with byzantine failures). We want to compute a path connecting two given vertices $s$ and $t$ which is as far away as possible from the obnoxious vertices. To be more precise, let's consider all the vertices $v(i)$ on the path between $s$ and $t$ and let $dmin(v(i))$ be the minimum distance from $v(i)$ to the closest obnoxious vertex. We want to find a path $P$ which maximizes the value $min\{dmin(v(i))|v(i) \text{ is a vertex on the path } P\}$.

First, we compute the values $dmin(i)$ for all the $n$ vertices of the graph. We set $dmin(ov(i)) = 0$ ($1 \leq i \leq k$) and $dmin(j) = +\infty$ (for $j \notin \{ov(1), \ldots, ov(k)\}$). We now run Dijkstra's algorithm, starting from multiple sources (i.e. $ov(1)$, ..., $ov(k)$). We will maintain a priority queue (e.g. binary heap, Fibonacci heap, set of buckets) $Q$ into which we insert all the obnoxious vertices (in the beginning). The key of every vertex $i$ inserted into $Q$ is $dmin(i)$. The rest of the algorithm is a normal implementation of Dijkstra's algorithm, i.e. when expanding a vertex $i$, we consider all the edges $(i, j)$ and, if $dmin(j) > dmin(i) + len(i, j)$, we set $dmin(j) = dmin(i) + len(i, j)$, delete $j$ from the priority queue (if it was previously inserted in it) and insert it again with the new value of $dmin(j)$. If all the $len(*, *)$ values are identical, we can consider $len(*, *) = 1$ and use BFS instead. We insert all the obnoxious vertices in the queue in the beginning (there are multiple vertices at distance 0) and then we repeatedly extract the vertex $i$ at the front of the queue, consider all the edges $(i, j)$, and if $dmin(j) > dmin(i) + 1$, we set $dmin(j) = dmin(i) + 1$ and insert $j$ at the end of the queue; initially, all the non-obnoxious vertices $j$ have $dmin(j) = +\infty$.

After computing the values $dmin(*)$, we want to find a path from $s$ to $t$ for which the minimum value $dmin(x)$ of a vertex $x$ on the path is maximum. We can reduce the problem to the maximum capacity path problem, discussed previously. We have two reduction possibilities. The first one consists of transforming the graph into a directed graph. We transform every vertex $u$ into two vertices $u_{in}$ and $u_{out}$ and add a directed edge $(u_{in}, u_{out})$ between them, whose capacity will be $dmin(u)$. Then, we transform each undirected edge $(u, v)$ into two directed edges, $(u_{out}, v_{in})$ and $(v_{out}, u_{in})$; the capacities of these edges will be $+\infty$. A maximum capacity path from $s_{in}$ to $t_{out}$ in the transformed graph is equivalent to a path from $s$ to $t$ which is farthest away from the obnoxious vertices in the initial graph. The second possible transformation is to maintain the graph undirected. We will add to each edge $(u, v)$ a capacity equal to $min\{dmin(u), dmin(v)\}$. Now, a maximum capacity path from $s$ to $t$ in this graph is farthest away from the obnoxious vertices in the initial graph.

If we want to compute a farthest distance tree connecting a source vertex $s$ to several destination vertices, we can use the same transformations and then use the solution for the maximum capacity tree problem presented earlier.

## 5. Farthest Distance Path Queries

Let's consider the same problem as in the previous section. We want to answer very fast queries of the following type: what is the path between two given vertices $s$ and $t$ which is farthest away from the obnoxious vertices ? Since running the algorithm presented in the previous section for each query would be too inefficient, we will focus on the second type of graph transformation we presented. With that transformation, the problem was reduced to finding a maximum capacity path between two given vertices. This problem was handled in a previous section, such that each query can be answered in a time complexity ranging from $O(log(n))$ to $O(1)$, after an $O(m \cdot log(m))$ time preprocessing.

## 6. Generalized Optimal Path Algorithm

We consider here a more general version of the problems presented in the previous section, in which every edge $(u, v)$ has $k$ non-negative weights $w_1(u, v)$, ..., $w_k(u, v)$. We also have $k$ aggregation functions, $f_1$, ..., $f_k$, from the set $\{min, max, +\}$. We want to compute the optimal paths from a set of source vertices $s_1$, ..., $s_m$ to every vertex in the graph. The weight of a path $P$ is a $k$-element array $(wP(1), \ldots, wP(k))$, where $wP(i)$ is the aggregate of all the weights $w_i(u, v)$ of the edges $(u, v)$ on $P$, using function $f_i$. We also have $k$ optimizaton functions $o_1, \ldots, o_k$, where $o_i = max$, if $f_i = min$, and $o_i = min$, if $f_i \in \{max, +\}$. A path $P_1$ is better than a path $P_2$ if there exists an index $q$ ($1 \leq q \leq k$), such that $wP_1(i) = wP_2(i)$ ($1 \leq i \leq q - 1$) and ($wP_1(q) \neq wP_2(q)$ and $o_q(wP_1(q), wP_2(q)) = wP_1(q)$). We can now use Dijkstra's algorithm on this graph. For each vertex $v$ of the graph, we will compute the $k$-element array $(wopt(v, 1), \ldots, wopt(v, k))$ corresponding to the best path from one of the source vertices to $v$. For a source vertex $s_i$, we have $wopt(s_i, j) = 0$, if $f_j \in \{max, +\}$ and $wopt(s_i, j) = +\infty$, if $f_j = min$. Since we have a total order over the set of all possible $k$-element arrays, every two arrays are comparable and, thus, we can correctly maintain the priority queue required by Dijkstra's algorithm. At the beginning, we insert all the source vertices in the priority queue. Then, as long as the queue is not empty, we extract the vertex $i$ with the smallest weight vector $(wopt(i, 1), \ldots, wopt(i, k))$ and "relax" its outgoing edges. For every such edge $(i, j)$, we check if the weight vector $wnew = (f_1(wopt(i, 1), w_1(i, j)), \ldots, f_k(wopt(i, k), w_k(i, j)))$ is better than the current weight vector of the vertex $j$; if it is, we set $wnew$ as the new weight vector of vertex $j$, remove $j$ from the priority queue and insert it back with the new weight vector. The time complexity of the generalized algorithm is the same as that of the original Dijkstra's algorithm, possibly multiplied by a factor of $k$ (if $k$ is not a constant). This generalization leads to many interesting problems, like, for instance, the *Maximum Capacity Shortest Path* problem. If we want to compute the path to a given destination vertex $t$ and $f_1 \in \{max, min\}$, we can binary search the weight of the first element and use the previously described algorithm on the remaining $k - 1$ weights, as a feasibility test (checking that a path exists if we ignore the edges $(u, v)$ with $w_1(u, v)$ smaller (larger) than a candidate value, for $o_1 = max(min)$). The time complexity of this approach is that of the optimal path computation algorithm, multiplied by $O(log(WMAX))$, where $WMAX$ denotes the range of the binary search for the first component of the weight vector (if we sort all the $w_1$ weights of all the edges and use this sorted array for the binary search, we have $WMAX = O(m)$, where $m$ is the number of edges of the graph).

## 7. Optimal k-Packet Routing with Ordering Constraints

Let's consider a network composed of $n$ vertices, $1, \ldots, n$. We want to send $k < n$ identical packets from the vertices $vinit(1), \ldots, vinit(k)$ (with $min\{vinit(1), \ldots, vinit(k)\} = 1$), such that at least one packet passes by every vertex, subject to the following constraint: after a packet is received by a vertex $i$, it can only be sent further to a vertex $j > i$. The cost of forwarding a packet from a vertex $i$ to a vertex $j > i$ is $c(i, j)$ and we want to minimize the total cost. The costs satisfy the triangle inequality, i.e. $c(i, j) \leq c(i, k) + c(k, j)$ (otherwise, we can replace the costs $c(i, j)$ by the length of the shortest path between the two vertices). We will compute $Cmin(v(1), \ldots, v(k))$, where $v(1) \geq \ldots \geq v(k)$ are the vertices at which the $k$ packets have currently arrived and all the vertices in the interval $[1, v(1)]$ have been visited by at least one packet. We initially have $Cmin(v_0(1), \ldots, v_0(k)) = 0$ (where $v_0(1), \ldots, v_0(k)$ is the descendingly sorted sequence of the values $vinit(1), \ldots, vinit(k))$ and $Cmin(*, \ldots, *) = +\infty$ for the other sequences. We will consider the sequences $(v(1), \ldots, v(k))$ which are lexicographically larger than the initial sequence $v_0(1), \ldots, v_0(k)$, in increasing lexicographic order. An $O(k \cdot n^{k+1})$ algorithm is easy to devise. For every sequence $v(1), \ldots, v(k)$, we consider every packet $j$ $(1 \leq j \leq k)$ and every previous vertex $v'(j) < v(j)$. For each such possibility, we sort the sequence $v(1), \ldots, v(j-1), v'(j), v(j+1), \ldots, v(k)$ decreasingly, as $v''(1) \geq \ldots \geq v''(k)$. Then, $Cmin(v(1), \ldots, v(k))$ is the minimum value among all the possibilities $Cmin(v''(1), \ldots, v''(k)) + c(v'(j), v(j))$. We can improve the time complexity to $O(k \cdot n^k)$, by always forwarding a packet to a vertex which was never visited by another packet. For every sequence, we consider the case of forwarding a packet to $v(1)$ from some vertex $v' < v(1)$. We have the following cases. If $v(1) - 1 > v(2)$, then $v' = v(1) - 1$ and $Cmin(v(1), \ldots, v(k)) = Cmin(v(1) - 1, v(2), \ldots, v(k)) + c(v(1) - 1, v(1))$. Otherwise, we consider every previous vertex $v' < v(1)$, sort decreasingly the sequence $v', v(2), \ldots, v(k)$, obtain a sequence $v''(1), \ldots, v''(k)$ and set $Cmin(v(1), \ldots, v(k)) = \min\{Cmin(v(1), \ldots, v(k)), Cmin(v''(1), \ldots, v''(k)) + c(v', v(1))\}$. There are $O(n^k)$ sequences for which we compute $Cmin(\ldots)$ in $O(k)$ time and $O(n^{k-1})$ sequences for which we compute $Cmin(\ldots)$ in $O(k \cdot n)$ time. The optimal cost is $\min\{Cmin(n, *, \ldots, *)\}$. From the $Cmin(*, \ldots, *)$ values we can easily derive the paths of the $k$ packets. Note that whenever we need to sort a sequence of vertices during the algorithm, this step can be performed in $O(k)$ time. This is because the new sequence is obtained from an already sorted sequence, where only one value is replaced; thus, in $O(k)$ time, we can find the (new) correct position of the new value.

Another version of this problem is the following: we have $n$ vertices and $m$ packet requests. The $i^{th}$ request asks that a packet is sent to vertex $r(i)$. We have $k$ packet flows, which need to satisfy the $m$ requests in order. Once a packet served a request $i$, we can send it to any vertex $r(j)$ $(j > i)$, in order to satisfy the request $j$. We maintain the packet transfer costs $c(a, b)$ between pairs of vertices $(a, b)$ of the graph. The dynamic programming formulation is as follows: we compute $Cmin(i, v(1), ..., v(k-1))$, with $v(q) \leq v(q+1)$ $(1 \leq q \leq k-2)$, meaning: the minimum cost of satisfying the first $i$ requests, such that $k-1$ packets are located at $v(1), \ldots, v(k-1)$ and the $k^{th}$ packet is located ar $r(i-1)$ (i.e. $v(k) = r(i-1)$). Initially, every packet $p$ $(1 \leq p \leq k)$ is located at the vertex $vinit(p)$. We will consider $r(0) = vinit(k)$ and have $Cmin(0, v'(1), \ldots, v'(k-1)) = 0$, where $v'(1), \ldots, v'(k-1)$ is the sorted sequence of the values $vinit(1), \ldots, vinit(k-1)$; we will have $Cmin(0, *) = +\infty$ for any other sequence of vertices. The dynamic programming equations are quite simple. We will initially consider that $Cmin(i \geq 1, *) = +\infty$. We consider the tuples $(i, v(1), \ldots, v(k-1))$ in increasing order of $i$ and, for each such tuple, we consider every packet $j$ to be sent to $r(i+1)$; thus, we will set $Cmin(i+1, v'(1), \ldots, v'(k-1)) = min\{Cmin(i+1, v'(1), \ldots, v'(k-1)), Cmin(i, v(1), \ldots, v(k-1)) + c(v(j), r(i+1))\}$, where $v'(1), \ldots, v'(k-1)$ is obtained by sorting in ascending order the sequence $v(1), \ldots, v(j-1), v(j+1), \ldots, v(k)$. In case the packet flows are different, we may drop the ordering constraint for the values $v(1), \ldots, v(k-1)$ of a state.

## 8. Optimal Path with Non-Linear Costs

Let's consider a path $v(1) = s, v(2), \ldots, v(n-1), v(n) = t$ from a source $s$ to a destination $t$ consisting of $n-2$ intermediate nodes. There are $k$ types of connections between any two consecutive nodes $v(i)$ and $v(i+1)$ $(1 \leq i \leq n-1)$. Each connection $(i, j)$ (of type $j$, between $v(i)$ and $v(i+1)$) has a latency of $l(i, j) \geq 0$ time units. When sending a packet from $s$ to $t$, we may use any of the $k$ connections at every step. Let's assume that the sum of the latencies of the connections of type $j$ traversed by the packet is $ltotal(j)$ $(1 \leq j \leq k)$. Then, the aggregate cost is $cagg = f(ltotal(1), \ldots, ltotal(k))$, where $f$ is a non-decreasing function on every argument. For instance, we may have $f(ltotal(1), \ldots, ltotal(k)) = g(c(1) \cdot ltotal(1)^{p(1)}, \ldots, c(k) \cdot ltotal(k)^{p(k)})$ $(c(j) > 0, p(j) \geq 1, 1 \leq j \leq k)$, where $g$ could be $+$ or $max$. We want to compute a packet sending strategy which minimizes the aggregate cost. When the latencies are integer numbers and the sums of the latencies of all the connections of the same type are not too large, we can use the following pseudo-polynomial dynamic programming algorithm. We will compute $Lmin(i, ltotal(1), \ldots, ltotal(k-1))$=the minimum sum of latencies of the type $k$ connections required to reach $v(i)$, if the sums of the latencies of the type $j$ connections $(1 \leq j \leq k-1)$ is $ltotal(j)$. We have $Lmin(1, 0, \ldots, 0) = 0$ and $Lmin(1, ltotal(1), \ldots, ltotal(k-1)) = +\infty$, if at least one value $ltotal(j)$ is larger than 0. For $i > 1$, we have: $Lmin(i, ltotal(1), \ldots, ltotal(k-1)) = min\{min\{Lmin(i, ltotal(1), \ldots, ltotal(j-1), ltotal(j) - l(i-1, j), ltotal(j+1), \ldots, ltotal(k-1)) \mid 1 \leq j \leq k-1\}, Lmin(i-1, ltotal(1), \ldots, ltotal(k-1)) + l(i-1, k)\}$. After computing these values, the minimum total cost is $min_{(ltotal(1), \ldots, ltotal(k-1))}\{f(ltotal(1), \ldots, ltotal(k-1), Lmin(n, ltotal(1), \ldots, ltotal(k-1)))\}$. For the case mentioned above, if $g = max$, we can also binary search the optimal aggregate cost on a suitable interval $[0, CMAX]$ with a suitable precision $\varepsilon > 0$. The feasibility test for a candidate value $Ccost$ consists of the following operations. For each type of connections $j$ we compute $lmax(j)$, the integer part of $(Ccost/c(j))^{1/p(j)}$. Then, we compute the same $Lmin$ values as before, except that the indices $ltotal(j)$ $(1 \leq j \leq k-1)$ are upper bounded by $lmax(j)$. If there exists at least a value $Lmin(n, ltotal(1), \ldots, ltotal(k-1)) < lmax(k)$, then $Ccost$ is a feasible value.

We can extend the algorithm to the case where the vertices form an arbitrary directed graph (not just a path) and we have multiple possible sources and destinations. In this case, we construct the expanded graph composed of tuples $(i, ltotal(1), \ldots, ltotal(k-1))$. We have an edge of cost 0 between a tuple $(i, ltotal(1), \ldots, ltotal(k-1))$ and a tuple $(j, ltotal(1), \ldots, ltotal(p-1), ltotal(p) + l(i, j, p), ltotal(p+1), \ldots, ltotal(k-1))$ if there exists a connection $(i, j)$ of type $p$ $(1 \leq p \leq k-1)$ in the original graph of latency $l(i, j, p)$. We also have edges of weight $l(i, j, k)$ between tuples $(i, ltotal(1), \ldots, ltotal(k-1))$ and $(j, ltotal(1), \ldots, ltotal(k-1))$, if there exists a type $k$ connection between $i$ and $j$, of latency $l(i, j, k)$. We now compute the shortest paths from any initial tuple $(s_i, 0, \ldots, 0)$ $(1 \leq i \leq$ number of sources$)$ to every tuple in the expanded graph (where $s_i$ is one of the source vertices in the original graph). In order to do this, the shortest path value corresponding to an initial tuple is set to 0 and all the initial tuples are inserted into the queue (or priority queue) used by the shortest path algorithm. This way, we compute the shortest path from any of the initial tuples using only one invocation of the shortest path algorithm.

Another solution consists of constructing the hyper-graph with vertices $(i, ltotal(1), \ldots, ltotal(k))$. We have an edge between a tuple $(i, ltotal(1), \ldots, ltotal(k))$ and a tuple $(j, ltotal(1), \ldots, ltotal(p-1), ltotal(p) + l(i, j, p), ltotal(p+1), \ldots, ltotal(k))$ if there exists a connection $(i, j)$ of type $p$ $(1 \leq p \leq k)$ in the original graph of latency $l(i, j, p)$. In this graph we only need to perform a breadth-first (BF) or depth-first traversal (DF) starting from each of the initial tuples and mark all the reachable states (in the case of BF, we do not insert into the queue an already marked state, and in the case of DF, we do not call the recursive traversal function for neighboring tuples that were previously marked). We can compute the non-dominated paths from the set of reachable states.

## 9. Optimal Paths with Multiple Objectives

We consider a directed graph with $n$ vertices and $m$ edges, where every edge $(i, j, eid)$, directed from $i$ to $j$, has $p$ costs: $c_q(i, j, eid) \geq 0$ ($1 \leq q \leq p$) ($eid$ distinguishes between multiple edges between the same pair of vertices and having the same orientation). Let $ne(i, j)$ be the number of edges directed from $i$ to $j$; these edges will be labeled $(i, j, eid)$ ($1 \leq eid \leq ne(i, j)$). For a path $P = v(1), \ldots, v(k)$ we can compute $p$ costs: $sc_{i|v(1),\ldots,v(k)} = f_i(psc_j = sc_{j|v(1),\ldots,v(k-1)}, \ldots (1 \leq j \leq p), pc_{l,eid} = c_l(v(k-1), v(k), eid), \ldots (1 \leq l \leq p))$, if the last edge of the path is $(v(k-1), v(k), eid)$. The functions $f_i$ can be arbitrary functions, e.g. $f_i = max$ (or $min$) $\{psc_i, pc_{i,eid}\}$, $f_i = psc_i + pc_{i,eid}$ / (if ($pc_{j,eid} = 0$) then $psc_j$ else $pc_{j,eid}$), or $f_i = psc_i + pc_{i,eid}$. When using the addition or the max functions, we will have $sc_{i|v(1)} = 0$, while for functions like $min$, we will have $sc_{i|v(1)} = +\infty$. We will also use a set of boolean comparison functions $better_j$, which are capable of comparing two values of the function $f_j$ and decide if the first one is better than the second one. We want to compute all the *non-dominated* paths from a set of source nodes $src_i$ to a set of destination nodes $dest_j$. A path $v(1), \ldots, v(k)$ is $non-dominated$ if there exists no other path $v'(1), \ldots, v'(k')$ (with $v'(1)$ being one of the source vertices and $v'(k')$ being one of the destination vertices), such that $better_j(sc_{j|v'(1),\ldots,v'(k')}, sc_{j|v(1),\ldots,v(k)}) = true$ for all the values of $j$ ($1 \leq j \leq p$).

We will present a solution for the case when every function has integer values, between 0 and $VMAX$. We will build a directed hyper-graph, in which every vertex is a tuple $(i, s_1, \ldots, s_p)$. We will add a directed edge from a state $(i, s_1, \ldots, s_p)$ to every state $(j, f_k(s_1, \ldots, s_p, c_l(i, j, eid)(1 \leq l \leq p))(1 \leq k \leq p))$, if the directed edge $(i, j, eid)$ exists in the original graph. We will now traverse the hyper-graph in a breadth-first manner starting from the vertices $(src_i, sc_{k|src_i}(1 \leq k \leq p))$ (we insert them all in the BF queue in the beginning) and mark all the reachable states. The non-dominated paths can then be computed easily.

If the functions $f_p$ are monotonically increasing (or decreasing), we can optimize the previous solution, as follows. Every state will be represented by a tuple $(i, s_1, \ldots, s_{p-1})$. For each such state we will compute a value $s_p(i, s_1, \ldots, s_{p-1})$ = the best $sc_p$ value of a path from one of the source nodes to node $i$, having the (other) costs $sc_1, \ldots, sc_{p-1}$ equal to $s_1, \ldots, s_{p-1}$. In order to compute these values, whenever we have a directed edge $(i, j, eid)$ in the original graph, we will add directed edges from a state $(i, s_1, \ldots, s_{p-1})$ to all the states $(j, f_k(s_1, \ldots, s_{p-1}, s_p(i, s_1, \ldots, s_{p-1}), c_l(i, j, eid)(1 \leq l \leq p))(1 \leq k \leq p-1))$. We will then compute a shortest path in this hyper-graph, starting from all the states $(src_i, sc_{k|src_i}(1 \leq k \leq p-1))$ (we add all of them in the beginning to the priority queue, and their initial costs will be $sc_{p|src_i}$). The path length optimization function will be $f_p$ (i.e. when expanding a tuple $(i, s_1, \ldots, s_{p-1})$, we set $sp(S) = min\{sp(S), f_p(s_1, \ldots, s_{p-1}, sp(i, s_1, \ldots, s_{p-1}), c_l(i, j, eid)(1 \leq l \leq p))\}$, where $S = (j, f_k(s_1, \ldots, s_{p-1}, s_p(i, s_1, \ldots, s_{p-1}), c_l(i, j, eid)(1 \leq l \leq p))(1 \leq k \leq p-1))$ and the directed edge $(i, j, eid)$ exists in the graph). Note that the graph is not fully known in the beginning. The edges between different states depend on the values $s_p(i, s_1, \ldots, s_{p-1})$ computed by the shortest path algorithm.

When the $f_i$ functions are addition functions and when we do not need to rebuild the non-dominated paths (we are only interested in their costs), we can improve the amount of memory we use. Let's consider $cmax_k = max\{c_k(i, j, eid)|(i, j, eid)$ is a directed edge in the graph$\}$. In order to compute $s_p(i, s_1, \ldots, s_{p-1})$ we only need values like $sp(j, s'_1 \geq s_1 - cmax_1, \ldots, s'_{p-1} \geq s_{p-1} - cmax_{p-1})$, and not all of the computed values. We can use only $O(cmax_1 \cdot VMAX^{p-2})$ memory for every node $i$ of the original graph. The value $sp(i, s_1, \ldots, s_{p-1})$ can be stored in an array $a(i)$ with $cmax_1$ entries, at the position $(s_1 \ mod \ cmax_1)$.

We can extend the problem as follows. We can use a directed edge $(i, j, eid)$ only if every function $f_k$, applied on the costs of a path from one of the source nodes to a node $i$, takes values from a given set $V(i, k)$. In this case, it is possible that, at some point, some of the edges may not be used. Because of this, we might need to insert zero, one or more "waiting edges" from a node $i$ to itself, which may have costs, but have no usage restrictions for the values of the functions $f_k$.

## 10. Maintaining Connectivity Information Under Edge Deletions

We are given an undirected graph with $n$ vertices and $m$ edges. We are also given a sequence of operations, consisting of queries and deletions. A deletion consists of deleting an existing edge $(i, j)$ from the graph. Each vertex $i$ of the graph has a weight $w(i)$. Each connected component $cc$ of the graph has a weight $wcc(cc)$, which is equal to the aggregate of the weights of the vertices contained in $cc$. The aggregate function is called $ccagg$ and could be, for instance, $+, max, min, *$ or any other commutative function. The graph itself also has a weight $wg$, which is computed as an aggregate over the weights of its connected components, using a commutative aggregate function $gagg$, for which an inverse function $gagg^{-1}$ exists (e.g. $+, *, xor$, etc.). We want to be able to efficiently support the following types of queries: 1) which is the weight $wg$ of the graph ? ; 2) which is the weight of the connected component containing vertex $x$ ? We will propose a solution based on the techniques developed for the *Union-Find* problem, when the entire sequence of queries and deletions is known in advance, i.e. in the offline case.

We will traverse the sequence of queries and deletions and perform all the deletions, without answering the queries. After all the deletions were performed, we compute the connected components of the graph, their weights, as well as the weight of the graph. We will represent every connected component as a rooted tree. At first, we choose an arbitrary vertex $x$ in each component and make $x$ the parent of all the other vertices in the component. The root of a tree contains the weight $wcc$ of the component. We now traverse the sequence of queries and deletions backwards (in reverse order). Whenever we encounter a deletion of an edge $(i, j)$, we insert the edge $(i, j)$ back into the graph. When doing this, we compute the two representatives $ri$ and $rj$ of the two vertices $i$ and $j$. If $ri \neq rj$, then we combine the connected components of the vertices $i$ and $j$ into one component. In order to do this, we will make one of the two representatives the parent of the other one. Assuming that we set $parent(ri) = rj$, we update the weight of the connected component: $wcc_{(new)}(rj) = ccagg(wcc_{(old)}(rj), wcc(ri))$. We also update the overall weight of the graph: $wg_{(new)} = gagg(gagg^{-1}(wg_{(old)}, wcc_{(old)}(rj)), wcc_{(new)}(rj))$ (e.g. $wg_{(new)} = wg_{(old)} - wcc_{(old)}(rj) + wcc_{(new)}(rj)$, for $gagg = +$). When we encounter a query for the connected component of a vertex $x$, we find the representative vertex $rx$ of the component containing vertex $x$ and set the query's result to $wcc(rx)$. When we encounter a query for the overall weight of the graph, we set the query's result to the current weight $wg$ of the graph. We use the union by rank heuristic for unions and the path compression technique when searching for the representative of the connected component of a vertex $x$. The problem can be extended by allowing vertex deletions. A vertex deletion consists of deleting all the adjacent edges and, afterwards, removing the remaining 1-vertex connected component from the graph. When traversing the sequence of operations backwards and we encounter a vertex deletion operation, we create a new connected component $cc$ containing the deleted vertex $x$, set its weight appropriately (possibly $wcc(cc) = w(x)$) and adjust the weight of the entire graph accordingly ($wg = gagg(wg, wcc(cc))$). Then, we insert back into the graph all the edges that were adjacent to $x$ and for which the vertex at the other endpoint exists in the graph. When inserting an edge $(x, y)$ back into the graph, we perform the same operations that we mentioned previously (when we inserted a deleted edge back into the graph).

## 11. Related Work

Computing optimal bottleneck paths and trees are fundamental problems which attracted a lot of attention both from a theoretical and a practical perspective. In [1], several assignment problems, including optimal bottleneck assignments, were studied. In [2], the problem of computing bottleneck multicast trees was studied and a modified Dijkstra's algorithm was proposed. In [3], the authors present the construction of a tree for answering bottleneck path queries, which is similar in essence to the solution we proposed for the maximum capacity path query problem. Efficient algorithms for constrained bottleneck paths and trees were also proposed in [11]. The problem of maintaining connectivity information in undirected graphs under an arbitrary sequence of edge insertions and deletions was studied in [6]. For a survey on dynamic graph algorithms, see [7]. In this paper we considered an easier version of the connectivity maintenance problem, where only edge deletions were allowed. Efficient packet routing and content delivery strategies in tree networks were considered in many papers (e.g. [8,9]). The optimal k-packet routing problem with ordering constraints is similar to the well-known bitonic tour problem for $k = 2$.

## 12. Conclusions

In this paper we addressed several fundamental theoretical problems with applications to the optimization of packet routing techniques. The problems spanned several important topics in Graph Theory and Network Optimization, like computing bottleneck paths (trees), computing optimal paths with non-linear costs, maintaining aggregate connectivity information under a sequence of network link failures (graph edge deletions) and computing optimal packet routing strategies. We presented new, efficient (and in several cases optimal) algorithms for all the problems we studied. Although the developed techniques assume that all the required information is available and stable, which is unrealistic for large scale distributed systems, the proposed algorithms can be used for performing local optimizations on some small parts of a distributed system. As future work, we intend to use some of the described algorithms and techniques for message routing optimization in a peer-to-peer communication framework.

Mugurel Ionuţ Andreica, Nicolae Ţăpuş
Department of Computer Science and Engineering
Politehnica University of Bucharest
Splaiul Independentei 313, sector 6, Bucharest, Romania
email:{*mugurel.andreica, nicolae.tapus*}*@cs.pub.ro*